\documentstyle[11pt,epsfig]{article}
\title{\bf The cosmological implications of a fundamental length:\\
a DSR inspired de-Sitter spacetime }
\author{N. Khosravi\thanks{email: n-khosravi@sbu.ac.ir}\,
and H. R. Sepangi\thanks{email: hr-sepangi@sbu.ac.ir}\\ {\small
Department of Physics, Shahid Beheshti University, Evin, Tehran
19839, Iran}}

\begin{document}
\maketitle
\begin{abstract}
We study a de-Sitter model in the framework of a Deformed Special
Relativity (DSR) inspired structure. The effects of this framework
appear as the existence of a fundamental length which influences
the behavior of the scale factor. We show that such a deformation
can either be used to control the unbounded growth of the scale
factor in the present accelerating phase or account for the
inflationary era in the early evolution of the universe.
\vspace{5mm}\noindent\\
PACS: 98.80.-k,04.60.Kz,04.20.-q
\end{abstract}

\maketitle
\section{Introduction}
It is generally believed that the existence of a fundamental
length is a natural feature in all theories that endeavor to
answer the old and interesting question of how to quantize gravity
\cite{gar,rovelli}. In string theory, considered as an alternative
to the unified theory of interactions, the fundamental length
plays a crucial role \cite{string}. Also, the fundamental length
appears in loop quantum gravity \cite{lqg} as a promising
alternative to quantum gravity, in the form of discrete spectra of
the area and volume operators \cite{thiemann}. This fundamental
length is introduced by hand in some effective theories
\cite{dsr,dsr2,ncgeometry} where it is shown that it can be
recovered as the limit of full quantum gravity
\cite{dsr,amelino,witten}.  Therefore, the motivation behind the
construction of these effective models is to study the effects of
such a fundamental length in simple scenarios which are amenable
to exact solutions. A common approach to introduce a fundamental
length is to modify or deform the algebraic structure of the
phase-space which may be done in various manners, a few example of
which can be found in \cite{noncomm,babak,nima}.

To study the effects of the existence of a fundamental length on the
behavior of the universe, one can construct a model based on the
noncommutative structure of DSR \cite{dsr} which is described by
what is known as the $\kappa$-deformation \cite{gilkman}. The
$\kappa$-deformation is introduced and studied in
\cite{ruegg,shahn}. The $\kappa$-Minkowski space
\cite{shahn,freidel} arises from the $\kappa$-Poincare algebra
\cite{ruegg} such that the ordinary brackets between coordinates are
replaced by
\begin{eqnarray}\label{kappa-nc}
\{{x}_0,{x}_i\}=\frac{1}{\kappa}{x}_i,
\end{eqnarray}
where $\{,\}$ represents the Poison bracket and $\kappa$ is the
deformation (noncommutative) parameter which has the dimension of
mass $\kappa=\epsilon \ell^{-1}$ for $c=\hbar=1$, where
$\epsilon=\pm1$ \cite{bruno} such that $\kappa$ and $\ell$ can be
interpreted as dimensional parameters corresponding to a fundamental
energy and length, respectively. This fundamental length can be
identified with an invariant minimum length {\it e.g.}, the Planck
length. In what follows, we restrict ourselves to the $\epsilon=1$
sector. In the next section we will study the phase-space structure
of a de-Sitter spacetime. The DSR inspired deformation in de-Sitter
spacetimes is introduced in  section \ref{3}. Sections \ref{4} and
\ref{5} deal with the interpretation of the results.

\section{The de-Sitter spacetime}
Let us start by briefly examine a simple de-Sitter model
\begin{eqnarray}\label{de-sitter}
ds^2=-N^2(t)dt^2+\frac{a^2(t)}{(1+\frac{k}{4}r)^2}(dx^2+dy^2+dz^2),
\end{eqnarray}
where $N(t)$ is the lapse function and $k=-1, 0 ,+1$ represents an
open, flat or closed universe respectively. The Einstein-Hilbert
Lagrangian with  cosmological constant $\Lambda$ becomes
\begin{eqnarray}\label{lagrangian}
{\cal{L}}&=&\sqrt{-g}(R[g]-2\Lambda)\nonumber\\
&=&-6N^{-1}a\dot{a}^2+6kNa-2\Lambda Na^3,
\end{eqnarray}
where $R[g]$ is the Ricci scalar and in the second line the total
derivative term has been ignored. The corresponding Hamiltonian, up
to a sign, becomes
\begin{eqnarray}\label{hamiltonian1}
{\cal{H}}_0&=&\frac{1}{24}Na^{-1}p_a^2+6kNa-2\Lambda Na^3.
\end{eqnarray}
Here, we note that since the momentum conjugate to $N(t)$,
$\pi=\frac{\partial{\cal{L}}}{\partial \dot{N}}$, vanishes, the
term $\lambda \pi$, where $\lambda$ is a Lagrange multiplier, must
be added as a constraint to Hamiltonian (\ref{hamiltonian1}), so
that the Dirac Hamiltonian is
\begin{eqnarray}\label{hamiltonian}
{\cal{H}}&=&\frac{1}{24}Na^{-1}p_a^2+6kNa-2\Lambda Na^3+\lambda\pi.
\end{eqnarray}
The equations of motion with respect to the above Hamiltonian become
\begin{eqnarray}\label{eqofmotion}
\dot{a}&=&\left\{a,{\cal{H}}\right\}=\frac{1}{12}Na^{-1}p_a,\\
\dot{p_a}&=&\left\{p_a,{\cal{H}}\right\}=\frac{1}{24}Na^{-2}p_a^2-6kN+6\Lambda
N a^2,\\
\dot{N}&=&\left\{N,{\cal{H}}\right\}=\lambda,\\
\dot{\pi}&=&\left\{\pi,{\cal{H}}\right\}=-\frac{1}{24}a^{-1}p_a^2-6ka+2\Lambda
a^3.\label{constraint}
\end{eqnarray}
Note that to satisfy the constraint $\pi=0$ at all times the
secondary constraint $\dot{\pi}=0$ must also be satisfied. A
simple calculations leads to
\begin{eqnarray}\label{eqofmotion1}
\dot{a}=\sqrt{\frac{1}{3}\Lambda a^2-k},\label{1equation}
\end{eqnarray}
where we have fixed the gauge by taking $N=1$, that is, we work in
the comoving gauge. Note that the other equations will be
satisfied automatically if the above equation is satisfied. The
solutions for non-vanishing $\Lambda$ become
\begin{eqnarray}\label{com.solutions}
a(t)=\frac{C_1}{4\Lambda}\left\{
\begin{array}{ll}
C_1^2e^{\sqrt{\frac{\Lambda}{3}}t}+24ke^{-\sqrt{\frac{\Lambda}{3}}t}\\
C_1^2e^{-\sqrt{\frac{\Lambda}{3}}t}+24ke^{\sqrt{\frac{\Lambda}{3}}t},
\end{array}\right.
\end{eqnarray}
where $C_1$ is the integration constant. For vanishing $\Lambda$ we
have
\begin{eqnarray}\label{com.solutions1}
a(t)=C_2\pm i\sqrt{k}t,
\end{eqnarray}
where $C_2$ is the integration constant and the solutions become
meaningful only for $k=0$ and $-1$. In all the possible solutions
the scale factor has a growing behavior without any limit, save
for the trivial cases $\Lambda=0$ and $k=0$. This means that the
scale factor goes to infinity for large $t$, either exponentially
or linearly for non-vanishing and vanishing $\Lambda$,
respectively.

\section{The DSR-inspired phase-space}\label{3}
It has long been argued that a deformation in phase-space can be
seen as an alternative path to quantization. This argument is
based on Wigner quasi-distribution function and Weyl
correspondence between quantum-mechanical operators in Hilbert
space and ordinary c-number functions in phase-space, see for
example \cite{quantization} and the references therein. The
deformation in the usual phase-space structure is introduced by
Moyal brackets which are based on the Moyal product
\cite{noncomm}. However, to introduce such deformations, it is
more convenient to work with Poisson brackets rather than Moyal
brackets.

From a cosmological point of view, models are built in a
minisuper-(phase)-space. It is therefore safe to say that studying
such a space in the presence of the deformations mentioned above
can be interpreted as studying the quantum effects on cosmological
solutions. One should note that in gravity and consequently in
cosmology, the effects of quantization are woven into the
existence of a fundamental length \cite{gar}, as mentioned in the
introduction. The question then arises as to what form of
deformations in phase-space is appropriate for studying quantum
effects in a cosmological model? The modified structure of the
geometry, that is, the noncommutative geometry \cite{ncgeometry},
has become the basis from which similar modifications in
phase-space have been inspired. In this approach, the fields and
their conjugate momenta play the role of the coordinate basis in
noncommutative geometry \cite{carmona}. Of course, in doing so an
effective model is constructed whose validity depends on its power
of prediction. For example, if in a model field theory the fields
are taken as noncommutative, as has been done in \cite{carmona},
the resulting effective theory predicts the same Lorentz violation
as a field theory in which the coordinates are considered as
noncommutative \cite{carlson}. Over the years, a large number of
works on noncommutative fields \cite{noncomm} have been inspired
by noncommutative geometry model theories \cite{ncgeometry}. As a
further example, it is well known that string theory can be used
to suggest a modification in the bracket structure of coordinates,
also known as Generalized Uncertainty Principe \cite{kempf}, which
is used to modify the phase-space structure \cite{babak}. In this
paper we will examine a new kind of modification in phase-space
inspired by relation (\ref{kappa-nc}), much the same as what has
been done in \cite{noncomm,babak,nima}.  In what follows we
introduce noncommutativity based on $\kappa$-Minkowskian space and
study its consequences on the solutions discussed in the previous
section.

To introduce noncommutativity we start from
\begin{eqnarray}\label{kappa-nc-filds}
\{N'(t),a'(t)\}= \ell a'(t),
\end{eqnarray}
where one can interpret $N(t)$ and $a(t)$, appearing as the
coefficients of $dt$ and $d\vec{x}$, in the same manner as ${x}_0$
and ${x}_i$ respectively. For this reason we name this kind of
phase-space, the $\kappa$-Minkowskian (minisuper) phase-space. For
the primed variables then, Hamiltonian (\ref{hamiltonian1})
becomes
\begin{eqnarray}\label{primedhamiltonian}
{\cal{H}}'_0&=&\frac{1}{24}N'a'^{-1}{p'}_{a}^2+6kN'a'-2\Lambda
N'{a'}^3,
\end{eqnarray}
where the ordinary Poisson brackets are satisfied except for
(\ref{kappa-nc-filds}). Following \cite{chai}, we introduce the
new variables
\begin{eqnarray}\label{newvariables}
\left\{
\begin{array}{ll}
N'(t)=N(t)-\ell a(t) p_a(t),\\
a'(t)=a(t).
\end{array}\right.
\end{eqnarray}
It can be easily checked that the above variables will satisfy
(\ref{kappa-nc-filds}) if the unprimed variables satisfy the
ordinary Poisson brackets. The term  $-\ell a(t) p_a(t)$ may be
looked upon as a direct consequence of a phase-space deformation
of relation (\ref{kappa-nc-filds}) which, as has been suggested,
could originate from string theory, noncommutative geometry and so
on, see \cite{nima,quantization,chern}. With the above
transformations Hamiltonian (\ref{primedhamiltonian}) changes to
\begin{eqnarray}\label{hamiltonianNC1}
{\cal{H}}^{nc}_0=\frac{1}{24}Na^{-1}{p}_{a}^2+6kNa-2\Lambda
N{a}^3-\frac{1}{24}\ell p_a^3-6\ell k a^2p_a+2\ell \Lambda a^4 p_a.
\end{eqnarray}
Clearly, the momentum $\pi$ conjugate to $N(t)$ does not appear in
(\ref{hamiltonianNC1}), {\it i.e.} it should be taken as the
primary constraint. It can be checked by using Legendre
transformations that the conjugate momentum corresponding to
$N(t)$, that is, $\pi=\frac{\partial{\cal{L}}}{\partial \dot{N}}$,
vanishes. Therefore, the term $\lambda\pi$ must be added to
Hamiltonian (\ref{hamiltonianNC1}), so that we find
\begin{eqnarray}\label{hamiltonianNC}
{\cal{H}}^{nc}=\frac{1}{24}Na^{-1}{p}_{a}^2+6kNa-2\Lambda
N{a}^3-\frac{1}{24}\ell p_a^3-6\ell k a^2p_a+2\ell \Lambda a^4
p_a+\lambda\pi.
\end{eqnarray}
The equations of motion with respect to Hamiltonian
(\ref{hamiltonianNC}) are
\begin{eqnarray}\label{NCeqofmotion}
\dot{a}&=&\left\{a,{\cal{H}}^{nc}\right\}=\frac{1}{12}Na^{-1}p_a-
\frac{1}{8}\ell p_a^2-6\ell k a^2
+2\ell\Lambda a^4,\\
\dot{p_a}&=&\left\{p_a,{\cal{H}}^{nc}\right\}=\frac{1}{24}Na^{-2}p_a^2-6kN
+6\Lambda
N a^2+12\ell kap_a-8\ell\Lambda a^3p_a,\\
\dot{N}&=&\left\{N,{\cal{H}}^{nc}\right\}=\lambda,\\
\dot{\pi}&=&\left\{\pi,{\cal{H}}^{nc}\right\}=-\frac{1}{24}a^{-1}p_a^2-6ka+2\Lambda
a^3.\label{NCconstraint}
\end{eqnarray}
Again we restrict ourselves to the comoving gauge for which $N=1$.
Combining the first and last equation we find
\begin{eqnarray}\label{equationNC}
\dot{a}+12\ell a^2\left(\frac{1}{3}\Lambda
a^2-k\right)=\sqrt{\frac{1}{3}\Lambda a^2-k},
\end{eqnarray}
which is compatible with other equations. Note that this equation
reduces to the commutative case (\ref{1equation}) when
$\ell\rightarrow 0$. The analytic solutions for this equation
exist only for the special cases $\Lambda=0$ or $k=0$. For
$\Lambda=0$, the solution of equation (\ref{equationNC}) is
complex
\begin{eqnarray}\label{lambda=0}
a(t)=\frac{(1+i)\tan[(1+i)\sqrt{6\ell}k^{\frac{3}{4}}(t+C_3)]}{2\sqrt{6\ell}k^{\frac{1}{4}}},
\end{eqnarray}
where $C_3$ is an integration constant. Real solutions are only
obtained for $k=0$ or $-1$, that is
\begin{eqnarray}\label{klambda=0}
a(t)&=&C_4,\hspace{4cm}k=0,\label{k=0NC}\\
a(t)&=&\frac{\tanh[2\sqrt{3\ell}(t+C_5)]}{2\sqrt{3\ell}},\hspace{1cm}k=-1,\label{k=-1}
\end{eqnarray}
where $C_4$ and $C_5$ are constants. The scale factor for $k=0$
and $\Lambda=0$, represented by equation (\ref{k=0NC}) is a
constant, similar to the commutative case. The solution for $k=-1$
is interesting since its behavior at late times, $t\rightarrow
\infty$, is completely different from that of the commutative
case. Here, the scale factor becomes constant, in contrast to the
commutative case given by equation (\ref{com.solutions1}) where
the scale factor grows linearly.

For the observationally interesting case $k=0$, the scale factor
can be written as
\begin{eqnarray}\label{k=0}
a(t)=3^{\frac{1}{6}}\left(\frac{e^{\sqrt{3\Lambda}\hspace{.5mm} t}
}{C_6+12\ell \sqrt{\Lambda} e^{\sqrt{3\Lambda}\hspace{.5mm} t}
}\right)^{\frac{1}{3}},
\end{eqnarray}
where $C_6$ is a constant of integration. This solution reduces to
the commutative one given by equation (\ref{com.solutions}) for
$\ell\rightarrow0$. Note that the above solution for $t\rightarrow
\infty$ becomes constant
\begin{eqnarray}
a(t\rightarrow\infty)=\left(\frac{\sqrt{3} }{12\ell
\sqrt{\Lambda}}\right)^{\frac{1}{3}},
\end{eqnarray}
which is different from that of the commutative solution given by
equation (\ref{com.solutions}). The general solution for the scale
factor given by equation (\ref{equationNC}) can not be obtained
analytically. This equation may however be solved numerically,
showing that these solutions follow the general behavior of the
special solutions, namely, the scale factor becomes constant for
$t\rightarrow \infty$ in contrast to the commutative case. This
behavior can be seen in figure 2 for different values of $k$.

The  scale factor calculated in (\ref{k=0}) and its behavior are the
main results of the present work. To describe these results we may
interpret $\Lambda$ both as a cosmological constant for the present
accelerating and the early inflationary phases of the universe. In
what follows we study the implications of the fundamental length on
the scale factor given by relation (\ref{k=0}) in these two distinct
scenarios.

\section{The present accelerating phase}\label{4}
The first scenario can be realized by noting that the main
ingredients in these calculations have been the introduction of
the above deformation to the ordinary Poisson brackets which
resulted in a damping exponential behavior in the evolution of the
universe. However, in contrast to usual scenarios where $\ell=0$,
the scale factor asymptotically approaches a constant value at
late times $t\rightarrow \infty$. It is clear from the plots in
figure 1 that for $\ell=0$ the curve is concave in all regions.
For $\ell\neq0$ the curve is concave at first but becomes convex
at late times\footnote{For the uninteresting case $\Lambda=0$ and
$k=-1$ the scale factor (\ref{k=-1}) is convex in all regions.}.
This means that the deceleration parameter is negative at the
beginning and becomes positive later\footnote{It appears that this
result is in contradiction with the observation that the
deceleration parameter is positive at early times and becomes
negative for late times, corresponding to a matter-radiation
dominated regime and a $\Lambda$ dominated phase, respectively
\cite{deceleration}. Here, our model is restricted to the
$\Lambda$ dominated phase and shows that the existence of $\ell$
can change the acceleration to a deceleration phase.}. Note that
the above results are a direct effect of the existence of a
non-vanishing fundamental length $\ell$ (noncommutativity
parameter). It is interesting to note that in studying the
phenomenological aspects of quantum gravity \cite{phn} one has to
go to high energies to zoom in on very small distances where we
expect the effects of quantum gravity to become dominant. However,
the results presented here point to an alternative possibility for
the observation of the fundamental length in that it is directly
related to the possibility of the observation of a constant scale
factor. The price to be paid however is that one has to wait,
possibly, for an incredibly long time. This could be interpreted
as the duality of time and energy, namely, either spending large
amounts of energy to penetrate small distances in a short span of
time or use little energy but face the possibility of having to
endure the passage of a long interval of time. In this scenario,
the cosmos can be taken as a large laboratory in which the study
of cosmological models can pave the way for a full understanding
of quantum gravity. Otherwise, ``an accelerator powerful enough to
study ... Planckian objects would have to be as large as the
entire galaxy'' \cite{susskind}.
\begin{figure}[th]
\centerline{\includegraphics[width=7cm]{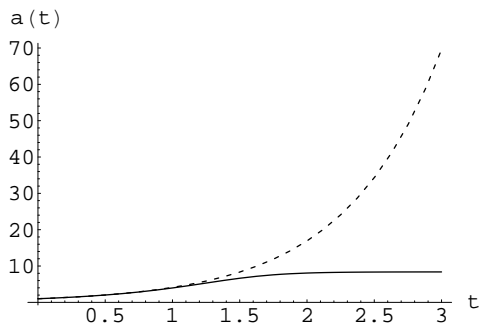}}
\caption{\label{fig}\footnotesize The solid line shows the scale
factor for non-vanishing $\ell$ ($\ell=0.0001$) and the dashed
line for vanishing $\ell$. The initial condition is $a(0)=1$ with
$\Lambda=6$, $k=0$ for both cases.}
\end{figure}
\begin{figure}
\begin{tabular}{ccc} \epsfig{figure=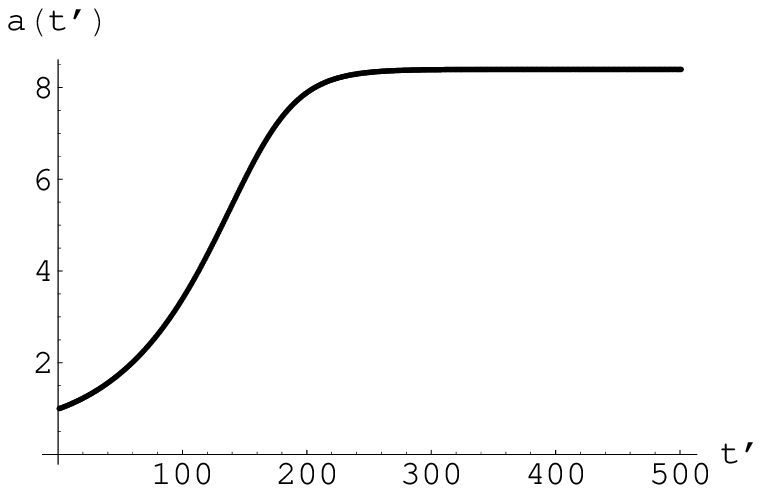,width=7cm}
\hspace{1cm} \epsfig{figure=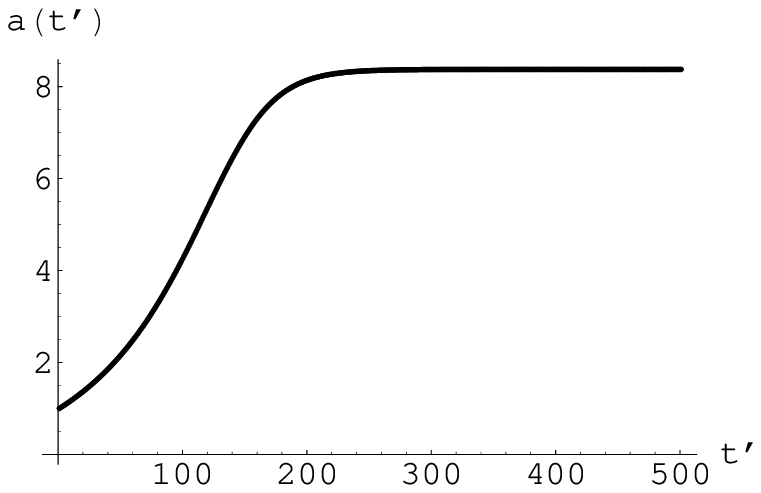,width=7cm}
\end{tabular}
\caption{\footnotesize The behavior of the scale factor for
$\Lambda=6$, $\ell=0.0001$,  $a(0)=1$ and $k=+1$, left, $k=-1$,
right. Note that we have re-scaled the horizontal axes such that
$t'=100t$.} \label{fig}
\end{figure}
\section{The early inflationary phase}\label{5}
The second scenario stems from the observation of the general
belief that the effects of a fundamental length or equivalently in
this model, the noncommutative structure, can be observed in the
early universe. It would therefore be interesting and natural to
discuss the results of the present work at the early stages of the
evolution of the  universe. We will show below that this simple
model may be useful in describing the early inflationary phase of
the universe. Let us then suppose that this phase begins at an
initial time $t_i$ and ends at a final time $t_R$, also known as
the reheating time \cite{brandenberger,linde}. During such a phase
the problem of e-folding, that is, the order of magnitude of the
size of the universe with which the scale factor expands during
inflation and the length of this period, which should be finite,
must be addressed whenever an inflationary model is discussed
\cite{inflation}. The value of the e-folding according to the
present data is at least $60$ which means that the scale factor of
the inflationary phase must satisfy the relation $60\sim \log
\frac{a(t_R)}{a(t_i)}$. From equation (\ref{k=0}), it is clear
that at early stages, since $\ell\ll1$, the scale factor behaves
exponentially and $\Lambda$ is viewed as the cosmological constant
of the inflationary era. As time passes, the behavior of the scale
factor would only depend on $\ell$ and clearly it would no longer
follow the exponential growth that it once had. This means that
the presence of $\ell$ could provide a natural mechanism with
which to exit from the inflationary phase. To address the
e-folding problem we note that one may introduce a relation
between the inflationary cosmological constant $\Lambda$ and the
fundamental length (noncommutativity parameter) $\ell$ using the
above definition for e-folding, that is, $\ell\sim\frac{e^{-180}
\sqrt{3}}{12\sqrt{\Lambda}}$ which would clearly address the $60$
e-folding problem \cite{inflation}. We may now see that such a
relatively simple model can address the above issues in
inflationary models. Note that the initial time can be chosen as
zero \cite{brandenberger}, as has been done in the above
calculations.

It would be appropriate here to point out that there are other
DSR-inspired theories for the inflationary era in the evolution of
the universe \cite{joao1}. These theories essentially amount to an
adaptation of the old varying speed of light idea to the more
compelling logics of the DSR setup, for a comprehensive review see
\cite{joao} and the references therein. In these models, the
assumption of a varying speed of light is invoked to address the
problems of the standard cosmology without resorting to the
ubiquitous scalar field invariably referred to as the inflaton.
The well known problems of the standard cosmology, namely the
flatness, cosmological constant, homogeneity and isotropy have all
a direct solution in such DSR-inspired models \cite{joao1}. An
important feature of these models is the functional form of $c(t)$
which, amongst other properties, should be able to account for the
rapid expansion of the universe during the inflationary era. To
this end, $c(t)$ is so constructed as to predict an almost
$10^{30}$ fold increase in the speed of light during inflation
over its present value. In contrast, the model presented in this
work has a deformation parameter, $\ell$, which makes a ``graceful
exit" from the inflationary era possible. The inflationary era
itself is produced by the existence of a cosmological constant,
$\Lambda$. In summary, in our model the existence of a
deformation, inspired by DSR, controls the rate of inflation. It
has also been shown \cite{nimaa} that in a DSR-inspired FRW model
the scale factor shows an exponential (inflationary) behavior just
after dust-domination as a consequence of the fundamental length.
This latter model is more relevant to other DSR-inspired works
\cite{joao1,joao} since in these models the inflationary behavior
appears just after the radiation-dominated era.

\section{Conclusions}\label{6}
More often than not, it is the case that the study of effective
theories can shed light on the blurred corners of the
corresponding full theory. The same is true in describing the
quantum effects on cosmology since the full theory is immensely
difficult to handle \cite{martin}. The DSR can be interpreted as
one such effective theory. Therefore, the study of cosmological
models within the framework of such effective theories could pave
the way for a more profound understanding. In the present study,
we have introduced a fundamental length by employing an effective
theory, namely a DSR-inspired model. In doing so we have chosen a
simple cosmological model, namely a de-Sitter spacetime. Here, the
introduction of a fundamental length causes additional terms to
appear in Hamiltonian (\ref{hamiltonianNC}) as compared to
(\ref{hamiltonian}). As has been mentioned in
\cite{nima,quantization,chern}, these extra terms can be
interpreted as the effects of high energy corrections of a full
theory, e.g., string theory\footnote{This interpretation is
consistent with the one introduced at the beginning of section
\ref{3} in that the extra terms can be interpreted as quantum
effects.}. Our results can also be used to address the problems of
interest in cosmology such as the inflation or present
accelerating phase, discussed above. It is therefore reasonable to
assume that the modifications introduced in this work, based on
relation (\ref{kappa-nc-filds}), are relevant in model theories
dealing with the problems mentioned above.

In summary, we have studied a noncommutative model theory in which
noncommutativity was inspired by invoking $\kappa$-deformation in
phase-space. The noncommutativity parameter appearing in our model
was interpreted as a fundamental length. Starting from a de-Sitter
universe, the effects of noncommutativity is either to predict a
``graceful exit'' from the inflationary phase or the emergence
from a late time accelerating phase. These results can be
interpreted as a phenomenological feature for the existence of
such a length and indirectly for the existence of a quantum theory
of gravity.
\vspace{8mm}\noindent\\
\textbf{Acknowledgments}\vspace{2mm}\noindent\\ We would like to
thank M. M. Sheikh-Jabbari for useful discussions. We also thank
M. Bojowald for useful comments and S. Sarkarati for his help in
preparing the plots.

\end{document}